%% file: main.tex
\documentclass[sigconf]{acmart}
\usepackage{multirow}
\usepackage{graphicx}
\usepackage{pifont}
\usepackage{algorithm}
\usepackage{algpseudocode}

\AtBeginDocument{%
  }
\usepackage{xpatch}

\makeatletter
\xpatchcmd{\ps@firstpagestyle}{Manuscript submitted to ACM}{}{\typeout{First patch succeeded}}{\typeout{first patch failed}}
\xpatchcmd{\ps@standardpagestyle}{Manuscript submitted to ACM}{}{\typeout{Second patch succeeded}}{\typeout{Second patch failed}}    \@ACM@manuscriptfalse
\makeatother

\renewcommand\footnotetextcopyrightpermission[1]{}
\setcopyright{none}

\begin{document}
\input{0_abs}

\title{PT-Mark: Invisible Watermarking for Text-to-image \\Diffusion Models via Semantic-aware Pivotal Tuning}

\author{Yaopeng Wang}
\authornotemark[2]
\affiliation{%
  \institution{School of Cyber Science and Engineering}
  \institution{Southeast University}
  \country{P. R. China}
}
\email{yaopengwang@seu.edu.cn}

\author{Huiyu Xu}
\authornotemark[2]
\affiliation{%
  \institution{The State Key Laboratory of Blockchain and Data Security}
  \institution{Zhejiang University}
  \country{P. R. China}
}
\email{huiyuxu@zju.edu.cn}

\author{Zhibo Wang}
\thanks{}
\affiliation{%
  \institution{The State Key Laboratory of Blockchain and Data Security}
  \institution{Zhejiang University}
  \country{P. R. China}
}
\email{zhibowang@zju.edu.cn}

\author{Jiacheng Du}
\affiliation{%
  \institution{The State Key Laboratory of Blockchain and Data Security}
  \institution{Zhejiang University}
  \country{P. R. China}
}
\email{jcdu@zju.edu.cn}

\author{Zhichao Li}
\affiliation{%
  \institution{The State Key Laboratory of Blockchain and Data Security}
  \institution{Zhejiang University}
  \country{P. R. China}
}
\email{mugen@zju.edu.cn}

\author{Yiming Li}
\affiliation{%
  \institution{Nanyang Technological University}
  \country{Singapore}
}
\email{ym.li@ntu.edu.sg}

\author{Qiu Wang}
\affiliation{%
  \institution{The State Key Laboratory of Blockchain and Data Security}
  \institution{Zhejiang University}
  \country{P. R. China}
}
\email{12321312@zju.edu.cn}

\author{Kui Ren}
\affiliation{%
  \institution{The State Key Laboratory of Blockchain and Data Security}
  \institution{Zhejiang University}
  \country{P. R. China}
}
\email{kuiren@zju.edu.cn}

\thanks{$\dagger$ Yaopeng Wang and Huiyu Xu are co-first authors.}

\maketitle

\input{1_intro}
\input{2_related}
\input{3_method}
\input{4_exp}

\input{6_conclusion}

\bibliographystyle{ACM-Reference-Format}
\bibliography{main}

\end{document}

%% file: 0_abs.tex
\begin{abstract}
Watermarking for diffusion images has drawn considerable attention due to the widespread use of text-to-image diffusion models and the increasing need for their copyright protection.
Recently, advanced watermarking techniques, such as Tree Ring, integrate watermarks by embedding traceable patterns~(e.g., Rings) into the latent distribution during the diffusion process. 
Such methods disrupt the original semantics of the generated images due to the inevitable distribution shift caused by the watermarks, thereby limiting their practicality, particularly in digital art creation.
In this work, we present Semantic-aware Pivotal Tuning Watermarks~(PT-Mark), a novel invisible watermarking method that preserves both the semantics of diffusion images and the traceability of the watermark. 
PT-Mark preserves the original semantics of the watermarked image by gradually aligning the generation trajectory with the original~(pivotal) trajectory while maintaining the traceable watermarks during whole diffusion denoising process. 
To achieve this, we first compute the salient regions of the watermark at each diffusion denoising step as a spatial prior to identify areas that can be aligned without disrupting the watermark pattern. 
Guided by the region, we then introduce an additional pivotal tuning branch that optimizes the text embedding to align the semantics while preserving the watermarks. 
Extensive evaluations demonstrate that PT-Mark can preserve the original semantics of the diffusion images while integrating robust watermarks. It achieves a 10\% improvement in the performance of semantic preservation~(i.e., SSIM, PSNR, and LPIPS) compared to state-of-the-art watermarking methods, while also showing comparable robustness against real-world perturbations and four times greater efficiency.
Moreover, PT-Mark can act as a plug-and-play module, aligning the generation process of existing advanced watermarking methods to make them more applicable in real-world scenarios.
\end{abstract}

%% file: 1_intro.tex
\section{Introduction}

Recently, the ease of use and powerful generation quality of text-to-image diffusion models~\cite{rombach2022high, saharia2022photorealistic, ramesh2022hierarchical, chang2023muse, podell2023sdxl} have led to a surge in their application across various fields, including digital art~\cite{huang2022draw, wang2024diffusion, podell2023sdxl} and filmmaking~\cite{ho2022video, xing2024survey, blattmann2023align, blattmann2023stable}. 
In these fields, images are generated based on carefully crafted text prompts provided by content creators, making these images highly valuable~\cite{rombach2022high, li2024towards, ramesh2022hierarchical, zhang2023adding}. Consequently, there is a growing need for copyright protection for images produced by diffusion models.



To effectively prove the ownership of the images generated by text-to-image diffusion models, many studies~\cite{wen2023tree, ci2024ringid, huang2024robin, zhang2024attack} have proposed embedding traceable watermark patterns into the initial noise~(latent state) of the latent diffusion process. Due to their low computational overhead and robustness against real-world perturbations~(e.g., JPEG compression), these watermarking methods have emerged as the prevailing solutions in both academic research and industrial applications.
They embed the watermark pattern into the Fourier transform of the initial noise and then perform the typical diffusion denoising process using the shifted initial noise to obtain the watermarked images.
For watermark verification, DDIM inversion is employed to recover the initial noise from the generated image and quantify the likelihood that the recovered watermark matches the target watermark.

\begin{figure*}
    \centering
    \includegraphics[width=1\linewidth]{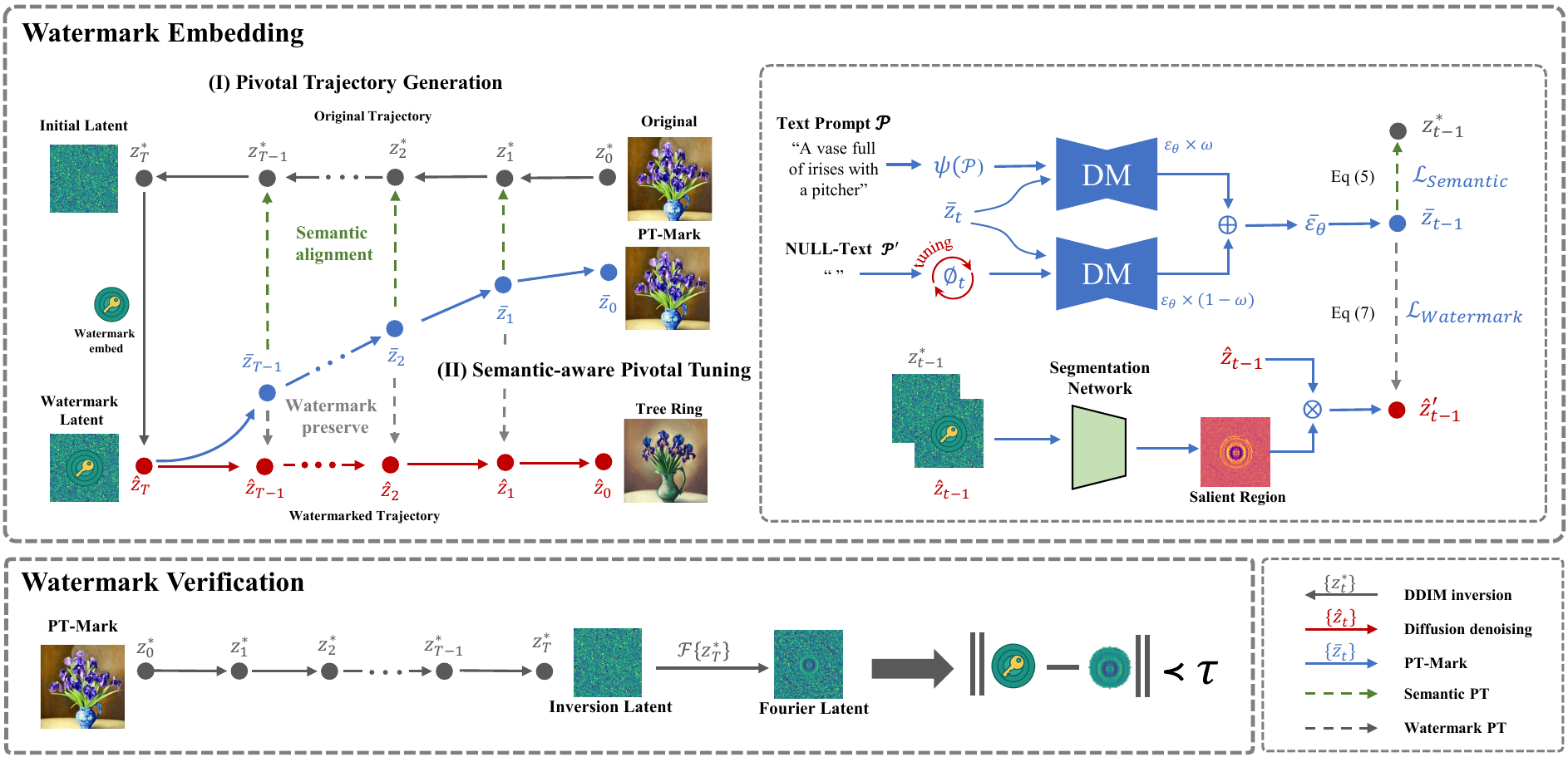}
    \vspace{-5mm}
    \caption{Overview of PT-Mark. We modify the entire diffusion denoising process after embedding watermarks into the initial latent $\hat{z_{T}}$ through Semantic-aware Pivotal Tuning. In this process, we optimize the null-text embedding to adjust the latent, making it closer to the original semantics~(Semantic PT) while preserving the watermark patterns~(Watermark PT).}
    \label{fig:overview}
\end{figure*}

Nevertheless, embedding watermarks inevitably induces a distributional shift in the initial noise, as shown in Figure~\ref{fig:overview}, leading to a deviation in the diffusion denoising process compared to the original generation trajectory.
This semantic drift reduces their usability in real-world scenarios, especially in domains where high semantic fidelity is essential, such as digital art creation~\cite{huang2022draw, wang2024diffusion, jiang2024cinematographic}.
To address this limitation, existing work~\cite{huang2024robin, zhang2024attack} focuses on optimizing the initial noise to find an alternative latent initialization that, when embedded with the watermark, generates a watermarked image that preserves the original semantics. 
Although these methods can somewhat mitigate the semantic drift caused by the initial distribution shift, they fail to effectively navigate the generation process, which involves multiple diffusion denoising steps primarily influenced by the text prompt. As a result, due to the lack of constraints on the generation process, these methods often still suffer from semantic degradation compared to the original semantics.

In this paper, we address the challenge of semantic preservation in watermarking methods for text-to-image diffusion models by considering the entire diffusion denoising process.
We argue that both the original diffusion trajectory and the watermarked trajectory can serve as explicit guidance, representing the semantics and the watermark, respectively.
By leveraging this guidance, we explicitly steer the entire diffusion denoising process through the manipulation of unconditional text embeddings, thereby enabling semantic control while preserving the traceability of the embedded watermark.
Building upon this basic idea, we propose Semantic-aware Pivotal Tuning Watermark~(PT-Mark), a novel invisible image watermarking method for diffusion models.
Specifically, for a given image to be watermarked, we first recover both the original generation trajectory and the watermarked trajectory by applying DDIM inversion and the typical watermarking process, respectively.
The full recovery of both trajectories enables us to explicitly capture the evolving semantic and watermark patterns at each timestep, as these patterns undergo changes throughout the diffusion denoising process.
We further propose employing a pretrained segmentation network to compare the latent states with and without the watermark, in order to identify the salient regions associated with the watermark. These regions serve as spatial priors to guide the disentanglement of semantic content from watermark patterns.
Next, we introduce a pivotal tuning branch that takes the learnable text embedding as input to generate a steering vector. Guided by the identified salient regions, the text embedding is optimized via two learning objectives: minimizing the discrepancy between the edited latent state and the original latent state within watermark-agnostic regions, and aligning it with the latent state in watermarked trajectory within watermark-intensive regions.

We validate the effectiveness of PT-Mark through extensive evaluations on two widely used datasets~(i.e., MS-COCO and Diffusion DB). Experimental results show that PT-Mark successfully preserves the original semantics, achieving a 10\% improvement in image quality metrics~(PSNR, SSIM, FID, and LPIPS) over state-of-the-art watermarking methods, while maintaining comparable robustness to real-world perturbations with a 99\% accuracy in watermark verification.
Moreover, PT-Mark offers a four times increase in efficiency compared to state-of-the-art methods and can be integrated as a plug-and-play module into current watermarking methods for text-to-image diffusion models.

In summary, our contributions are three-fold:
\begin{itemize}
    \item 
    We propose Semantic-aware Pivotal Tuning Watermark~(PT-Mark), which explicitly edits the entire diffusion denoising process by optimizing the text embedding to disentangle semantics from watermark patterns, thereby aligning image semantics while preserving watermark traceability.
    \item PT-Mark can act as a plug-and-play module, integrating into existing watermarking methods for text-to-image diffusion models to effectively refine the generation trajectory without compromising the embedded watermark. 
    \item Extensive experiments demonstrate that PT-Mark surpasses state-of-the-art watermarking methods in both invisibility~(achieving a 10\% improvement in PSNR) and efficiency~(reducing generation time by a factor of four), while maintaining high watermark extraction accuracy (>99\%) under diverse real-world perturbations.
\end{itemize}

%% file: 2_related.tex
\section{Related Work and Background}
\label{sec:preliminary}
\noindent\textbf{Diffusion Models and DDIM Inversion. } 
Diffusion models have demonstrated remarkable success in high-fidelity image generation and have become foundational for a wide range of generative tasks, including text-to-image synthesis~\cite{avrahami2023spatext, nichol2021glide, rombach2022high, podell2023sdxl}, image editing~\cite{hertz2022prompt, cho2024noise, mokady2023null, chen2024tino}, and restoration~\cite{xia2023diffir, yue2024efficient}. These models progressively transform random noise into coherent images through a learned denoising process, effectively modeling complex data distributions. A representative framework is the Denoising Diffusion Probabilistic Model~(DDPM)~\cite{ho2020denoising}, which gradually adds Gaussian noise to the data sample across $T$ timestamps according to a predefined noise schedule $\{\beta_{t}\}$. The forward process is defined as:
\begin{equation}
\label{eq:diffusion}
    z_{t}=\sqrt{\bar{\alpha}_{t}} z_{0}+\sqrt{1-\bar{\alpha}_{t}} \epsilon,
\end{equation}
where $\epsilon \sim \mathcal{N}(0, \mathbf{I})$ is Gaussian noise, and $\bar{\alpha}_t=\prod_{i=1}^t(1-\beta_i)$ is derived from a predefined noise schedule $\{\beta_{t}\}$. Starting from pure noise $z_{T}$ , the reverse process iteratively removes the noise to recover the data sample $z_{0}$. 
To improve sampling efficiency, Denoising Diffusion Implicit Models~(DDIM)~\cite{song2020denoising} reformulate the reverse process into a deterministic and non-Markovian process. Rather than sampling from conditional distributions, DDIM directly predicts the previous latent state $z_{t-1}$ using:
\begin{equation}
\label{eq:ddim denoising}
    z_{t-1}=\sqrt{\alpha_{t-1}}\left(\frac{z_t-\sqrt{1-\alpha_t}\epsilon_\theta(z_t)}{\sqrt{\alpha_t}}\right)+\sqrt{1-\alpha_{t-1}}\epsilon_\theta(z_t).
\end{equation}
This inversion mechanism enables recovery of latent noise while maintaining consistency with the generation process, facilitating downstream tasks such as image editing.


\vspace{0.3em}
\noindent\textbf{Null-text Inversion.}\label{sec:null-text}  Classifier-Free Guidance~(CFG)~\cite{ho2022classifier} is widely utilized in diffusion models for conditional text-driven image generation. It enhances the generated images' fidelity to textual condition $\mathcal{P}$ by combining conditional and unconditional predictions during the reverse diffusion process. Specifically, CFG employs a null-text embedding $\emptyset$ to compute unconditional predictions and amplifies conditional predictions based on a guidance strength parameter:
\begin{equation}
\label{eq:cfg}
    \tilde{\epsilon}_\theta(z_t,t,\mathcal{C},\emptyset)=w\cdot\epsilon_\theta(z_t,t,\mathcal{C})+(1-w)\cdot\epsilon_\theta(z_t,t,\emptyset).
\end{equation}
Here, $z_t$ represents latent variables at time step $t$, $\mathcal{C}=\psi(\mathcal{P})$ is the embedding of the text condition, and $w$, typically greater than 1, controls the strength of the conditioning. However, using large guidance scales in CFG can lead to cumulative errors, reducing both image fidelity and editability.
Null-text Inversion~(NTI)~\cite{mokady2023null} addresses these challenges by dynamically optimizing the null-text embedding during the reverse diffusion process of diffusion models. Initially, NTI performs DDIM inversion with a guidance scale $w=1$, generating a sequence of reference noise maps $\{z_t^*\}_{t=1}^T$. Subsequently, the null-text embedding $\emptyset_{t}$ is optimized at each diffusion step to minimize the reconstruction error $||z_{t-1}-z_{t-1}^*||_2^2$.

\noindent\textbf{Image Watermarking for Diffusion Models. }Image watermarking aims to embed identifiable information into images to protect intellectual property and trace content origin. It has evolved into two main approaches for generative models: embedding watermarks after image generation~(post-hoc) or during the generation process~(in-generation). Post-hoc methods, such as those using frequency domain transformations like Discrete Wavelet Transform (DWT)~\cite{xia1998wavelet} and Discrete Cosine Transform~(DCT)~\cite{cox2007digital}, or encoder-decoder architectures like HiDDeN~\cite{zhu2018hidden} and StegaStamp~\cite{tancik2020stegastamp}, embed watermarks into pre-generated images. RivaGAN~\cite{zhang2019robust} introduced adversarial training to enhance robustness, though these methods still face challenges balancing watermark strength and image quality. In-generation watermarking modifies the generation process itself, embedding the watermark during model training or in the latent space. Stable Signature~\cite{fernandez2023stable} fine-tunes the latent decoder to embed watermarks without altering the image generation pipeline significantly. However, this method requires training a separate decoder for each user to ensure unique watermarks, which limits its scalability and flexibility. Tree-Ring watermarking~\cite{wen2023tree} offers a more integrated approach by embedding watermarks through modifications to the initial noise distribution. However, this approach alters the latent distribution, potentially causing semantic changes and compromising the fidelity of the original model. Recent works like Zodiac~\cite{zhang2024attack} and ROBIN~\cite{huang2024robin} aim to improve image quality while maintaining watermark robustness. However, these methods still face challenges such as color shifts and the introduction of artifacts, which degrade the visual quality. 
In response, we propose PT-Mark, a watermarking method designed to effectively balance robustness and invisibility.

%% file: 3_method.tex
\section{PT-Mark}
In this section, we present our method, PT-Mark, which explicitly modifies the diffusion denoising process to realign the semantics disrupted by watermarks while preserving their traceability.
We begin by formulating the problem of invisible watermarking in Section~\ref{sec:problem_forumulation}, followed by an overview of our method in Section~\ref{sec:overview}. Finally, we detail the key components of PT-Mark in Section~\ref{sec:design_details}.


\subsection{Problem Formulation}
\label{sec:problem_forumulation}

\noindent\textbf{Watermark embedding}. 
Given an input image $x$ to be watermarked, the watermark embedder first estimates its generation trajectory $z^{*}$ via DDIM inversion. Subsequently, a specific watermark message $W$, such as the ring radius used in Tree-Ring~\cite{wen2023tree} to uniquely identify the user, is embedded into the initial noise $z_{T}^{*}$ within a user-specific region $\mathbb{M}$, where $\mathbb{M}$ is a binary mask indicating the spatial location of the watermark. The modified initial noise, denoted as $\hat{z}_{T}$, serves as the new starting point for the generation trajectory of the watermarked image. Guided by a user-provided text prompt, the watermark embedder employs the diffusion model to iteratively denoise $\hat{z}_{T}$ through a sequence of diffusion denoising steps, producing a latent trajectory $\hat{z}_{T} \rightarrow \hat{z}_{0}$. Once the final latent state $\hat{z}_{0}$ is obtained, it is passed through the decoder to generate the final watermarked image $\hat{x}$.


\vspace{0.3em}
\noindent\textbf{Watermark verification}. 
Given an watermarked image $\hat{x}$, the watermark verification process employs DDIM inversion to estimate the original generation trajectory $\{z^{*}_{t}\}_{t=0}^T$. The verifier then performs a statistical test to compute the p-value, which quantifies the likelihood that the observed watermark could appear in a natural image by random chance. To compute the p-value, the verifier first transforms $z_{T}^{*}$ into the Fourier domain, yielding the representation $y$. The null hypothesis is formulated as $H_{0}: y \sim \mathcal{N}(0,\sigma^{2},I_{C})$, where $\sigma^{2}$ is estimated per image based on the variance of $y$, reflecting the fact that DDIM inversion maps any input image to a Gaussian-distributed noise space. To evaluate this hypothesis, the verifier computes a distance score that quantifies the discrepancy between the embedded watermark message $W$ and the extracted message $y$ within the region specified by the binary mask $\mathbb{M}$.
\begin{equation}
    \eta = \frac{1}{\sigma^{2}}\sum(\mathbb{M} \odot W - \mathbb{M} \odot y)^{2}.
\end{equation}
An image is classified as watermarked if the computed value of $\eta$ is sufficiently small to be unlikely under random chance. The corresponding p-value, representing the probability of observing a value less than or equal to $\eta$, is computed from the cumulative distribution function~(CDF) of the non-central chi-squared distribution. Typically, non-watermarked images exhibit higher p-values, whereas watermarked images produce lower p-values. A sufficiently low p-value leads to the rejection of the null hypothesis $H_{0}$, thereby confirming the presence of the watermark.

\vspace{0.3em}
\noindent\textbf{Invisible Watermarking}. 
In the watermark embedding process, the original initial noise $z^{*}_T$ is modified into a new latent state $\hat{z}_T$ containing the watermark. This modified initial noise $\hat{z}_{T}$ serves as the new starting point for the generation trajectory ${\hat{z}_T \rightarrow \hat{z}_0}$. However, due to the modification in the initial state, the resulting generation trajectory ${\hat{z}_T \rightarrow \hat{z}_0}$ deviates from the original trajectory ${z^{*}_T \rightarrow z^{*}_0}$, leading to semantic inconsistencies between the generated watermarked image $\hat{x}$ and the original image $x$. Therefore, the primary objective of invisible watermarking is to preserve the semantic consistency between $\hat{x}$ and $x$. This can be measured by comparing the feature distances, such as the FID score: $FID(\hat{x}) \simeq FID(x)$.
At the same time, invisible watermarking need to ensure that the watermark’s traceability is preserved, meaning the DDIM trajectory of $\hat{x}$, denoted as $\{\hat{z}_0,\dots,\hat{z}_{T}\}$, should be close to the original trajectory $\{z^{*}_0,\dots,z^{*}_{T}\}$.




\subsection{Overview}
\label{sec:overview}
The high-level overview of PT-Mark is illustrated in Figure~\ref{fig:overview}. The core idea of our method is to explicitly learn the disentanglement of semantics content and watermark patterns in the latent space by optimizing the null-text embedding to guide the diffusion denoising process. This design enables the realignment of the generation trajectory with the original semantics while preserving the traceability of the embedded watermark.

PT-Mark consists two main stages in watermark embedding: Pivotal Trajectory Generation and Semantic-aware Pivotal Tuning. 
In the Pivotal Trajectory Generation stage, we generate both the original and watermarked trajectories to serve as learning pivots for semantic content and watermark information, respectively. Specifically, we leverage DDIM inversion to recover the original trajectory, and execute a typical watermarked image generation process to obtain the watermarked trajectory. By comparing the latent states along these two trajectories, we compute a salience map that explicitly segments the watermark-relevant information at each diffusion step. 
In the Semantic-aware Pivotal Tuning stage, we introduce an additional pivotal tuning branch, referred to as the PT-Mark trajectory. In this trajectory, we optimize a null-text embedding~(as described in Section~\ref{sec:preliminary}), and input it into the diffusion model at each step to generate a steering vector. This steering vector adjusts the latent states along the PT-Mark trajectory, promoting the disentanglement of semantic content and watermark patterns. It ensures that the edited latent state closely aligns with the original trajectory, while preserving the salient features of the watermark within the watermarked latent space.

For watermark verification, we employ DDIM inversion to estimate the initial latent variable $z_{T}^{*}$ and subsequently transform it into the Fourier domain for comparison against the watermark message stored in the database. As PT-Mark adopts the same verification procedure as existing advanced diffusion-based watermarking methods~\cite{huang2024robin, wen2023tree, zhang2024attack, ci2024ringid}, it functions as a plug-and-play solution. Notably, PT-Mark operates by modifying only the diffusion denoising process, without requiring additional training of the diffusion model.
This allows PT-Mark to be easily integrated into existing watermarking methods for text-to-image diffusion models.

\subsection{Design Details}
\label{sec:design_details}
\textbf{Pivotal Trajectory Generation}.
To effectively identify the semantic and watermark distribution in latent space, facilitating the learning of the decoupling between watermark and semantics in subsequent stages, it is crucial to segment a reliable spatial guidance. To this end, we first employ DDIM inversion with a guidance scale of $w = 1$ to provide a rough approximation of the original image and enable us to recover the original trajectory $z_{0}^{*}\rightarrow z_{T}^{*}$. In parallel, we embed the watermark information into the initial latent $z_{T}^{*}$ to obtain the modified latent $\hat{z}_{T}$, which then serves as the starting point for the diffusion denoising process to generate the watermarked trajectory $\hat{z}_{T}\rightarrow \hat{z}_{0}$. At each timestep along these two trajectories, we extract the latent states and input them into a pretrained segmentation network $f_{seg}$, which segments the salient regions by highlighting the differences between the original and watermarked latents. This yields a mask that localizes the watermark-relevant areas, which then serves as reliable guidance for subsequent disentanglement optimization.


\vspace{0.3em}
\noindent \textbf{Semantic-aware Pivotal Tuning}.
We initiate the editing of the entire diffusion denoising process from the initial noise of the watermarked trajectory. The core objective of this editing is to manipulate the semantics such that they closely align with the original, while simultaneously preserving the embedded watermark. Achieving this goal requires effective disentanglement of semantic content and watermark information, which is challenging when directly modifying the latent states. Inspired by the Null-text Inversion technique introduced in Section~\ref{sec:preliminary}, we formulate this editing as a generative optimization problem, implemented through an additional tuning branch referred to as the PT-Mark trajectory. Leveraging the expressive power of text embeddings in the latent diffusion process, which inherently guide generation and influence its trajectory, we introduce an optimizable null-text embedding $\emptyset$. This embedding is iteratively optimized to capture the fine-grained differences between the original and watermarked trajectories, thereby enabling effective disentanglement of semantics and watermark patterns. 


Specifically, given a text prompt $\mathcal{P}$, we first encode it into the textual condition embedding $\mathcal{C} = \psi(\mathcal{P})$, and initialize the optimizable null-text embedding as empty, denoted by $\emptyset_{T}=\psi("\ ")$. These two embeddings are then combined to generate a steering vector that guides the modification of the latent state at each step of the diffusion denoising process. At each timestamp $t$, the null-text embedding $\emptyset_{t}$ is initialized with the embedding from the previous step, $\emptyset_{t+1}$. Leveraging the original and watermarked trajectories generated by the Pivotal Trajectory Generation stage, we optimize the null-text embedding $\emptyset$ with the default guidance scale $w=7.5$ across the timesteps $t=T,\dots,1$, each for $N$ iterations. This optimization process is driven by two objectives (i).~\textbf{Semantic maintenance}: we optimize the null-text embedding to ensure that the edited latent state remains close to the corresponding latent in the original trajectory $\{z^{*}_{t}\}_{t=1}^T$: 
\begin{equation}
\label{eq:Semantic}
    \mathcal{L}_{Semantic}=\left\|z_{t-1}^{*}-z_{t-1}(\bar{z}_{t},\emptyset_{t},\mathcal{C})\right\|_{2}^{2}.
\end{equation}
~(ii).~\textbf{Watermark preservation}: Guided by the spatial prior obtained from the Pivotal Trajectory Generation stage~(in the form of a spatial mask $M$), we simultaneously optimize the null-text embedding to constrain the edited latent state to remain aligned with the watermarked trajectory $\{\hat{z}_{t}\}_{t=T}^1$, thereby preventing the loss of watermark information during the diffusion denoising process, specifically within the salient regions defined by $M$:
\begin{equation}
\label{eq:mask}
    M = f_{seg}(\hat{z}_{t-1}, {z}^{*}_{t-1}),
\end{equation}
\begin{equation}
\label{eq:Watermark}
    \mathcal{L}_{Watermark}=\left\|M \odot \{\hat{z}_{t-1}- z_{t-1}(\bar{z}_{t},\emptyset_{t},\mathcal{C})\}\right\|_{1},
\end{equation}
where $z_{t-1}$ is a temporary variable during the optimization over $N$ iterations. At the end of each step $t$, we update the latent for the next step as:
\begin{equation}
\label{eq:update}
    \bar{z}_{t-1}=z_{t-1}(\bar{z}_t,\emptyset_t,\mathcal{C}).
\end{equation}

By integrating the two optimization objectives, we continuously optimize the null-text embedding to ultimately generate the final watermarked image. The overall loss function is formulated as:
\begin{equation}
\label{eq:all}
    \mathcal{L} = \lambda_{1}\mathcal{L}_{Semantic}+\lambda_{2}\mathcal{L}_{Watermark},
\end{equation}
where $\lambda_{1},\lambda_{2}$ are the weight that balance the semantic alignment and watermark preservation objectives, respectively.
This process is repeated until the image semantics are sufficiently close to the original image. The pseudo code for Semantic-aware Pivotal Tuning is presented in Alg.~\ref{alg:mark}.
\begin{algorithm}[ht]
\caption{Semantic-aware Pivotal Tuning}\label{alg:mark}
\begin{algorithmic}[1]
\State \textbf{Input:} source prompt $\mathcal{P}$, input image $x$, pre-trained segmentation network $f_{seg}$.
\State \textbf{Output:} Noise vector $\bar{z}_T$ and optimized embeddings $\{\emptyset_t\}_{t=1}^T$.
\State Set guidance scale $w = 1$;
\State Compute the original trajectory $\{z^{*}_{t}\}_{t=1}^T$ using DDIM inversion over $x$;
\State Set $\hat{z}_{T} \leftarrow Embed(z^{*}_{T})$ \quad $\rhd$ Embed watermarking into $z^{*}_{T}$;
\State Compute the watermarked trajectory $\{\hat{z}_{t}\}_{t=T}^1$ using DDIM sampling initialized from $\hat{z}_{T}$;
\State Set guidance scale $w = 7.5$;
\State Initialize $\bar{z}_{T} \leftarrow \hat{z}_{T}, \mathcal{C} \leftarrow \psi(\mathcal{P}), \emptyset_T \leftarrow \psi("\ ")$;
\For{$t = T, T-1, \ldots, 1$}
    \For{$j = 0, \ldots, N-1$}
        \State $M=f_{seg}(\hat{z}_{t-1},z_{t-1}^*)$;
        \State $\mathcal{L}_{Semantic} \leftarrow \left\|z_{t-1}^{*}-z_{t-1}(\bar{z}_{t},\emptyset_{t},\mathcal{C})\right\|_{2}^{2}$;
        \State $\mathcal{L}_{Watermark} \leftarrow \left\|M \odot \{\hat{z}_{t-1}- z_{t-1}(\bar{z}_{t},\emptyset_{t},\mathcal{C})\}\right\|_{1}$;
        \State $\emptyset_t \leftarrow \emptyset_t - \nabla_{\emptyset}(\lambda_{sem}\mathcal{L}_{Semantic}+\lambda_{wm}\mathcal{L}_{Watermark})$;
    \EndFor
    \State Set $\bar{z}_{t-1} \leftarrow z_{t-1}(\bar{z}_t, \emptyset_t, \mathcal{C})$, $\emptyset_{t-1} \leftarrow \emptyset_t$;
\EndFor
\State \textbf{Return} $\bar{z}_T$, $\{\emptyset_t\}_{t=1}^T$
\end{algorithmic}
\end{algorithm}


%% file: 4_exp.tex
\begin{table*}[]
\centering
\caption{The quantitative evaluation results on Diffusion DB and MS-COCO. Best results are in \textbf{bold}.}
\label{tab:quantitative}
\resizebox{0.98\textwidth}{!}{%
\begin{tabular}{l|ccccc|cccccccc}
\hline
\multicolumn{1}{c|}{\multirow{2}{*}{Method}} & \multicolumn{5}{c|}{Image Quality} & \multicolumn{7}{c}{Watermarking Robustness}    &      \\
\multicolumn{1}{c|}{} &
  PSNR $\uparrow$ &
  SSIM $\uparrow$ &
  MSSIM $\uparrow$ &
  FID $\downarrow$ &
  LPIPS $\downarrow$ &
  Clean &
  JPEG &
  Crop &
  Blur &
  Noise &
  Bright &
  Rotation &
  Avg \\ 
\hline
\multicolumn{14}{c}{\textit{Diffsuion DB}} \\ 
\hline
DwtDct                                        & 38.20 & 0.97 & \textbf{0.99} & \textbf{1.28}  & \textbf{0.01} & 0.86 & 0.51 & 0.54 & 0.49 & 0.44 & 0.49 & 0.49 & 0.49 \\
DwtDctSvd                                     & 38.19 & \textbf{0.98} & \textbf{0.99} & 4.00  & \textbf{0.01} & \textbf{1.00} & 0.52 & 0.50 & 0.78 & 0.58 & 0.48 & 0.44 & 0.55 \\
RivaGAN                                       & \textbf{40.52} & \textbf{0.98} & \textbf{0.99} & 6.14  & \textbf{0.01} & 0.99 & 0.82 & 0.97 & 0.82 & 0.70 & 0.86 & 0.43 & 0.77 \\
StegaStamp                                    & 28.53 & 0.91 & 0.94 & 24.88 & 0.03 & \textbf{1.00} & \textbf{1.00} & 0.62 & 0.95 & 0.83 & 0.89 & 0.45 & 0.79 \\ \hline
Tree-ring                                    & 15.18 & 0.56 & 0.59 & 42.97 & 0.37 & \textbf{1.00} & \textbf{1.00} & \textbf{1.00} & \textbf{1.00} & \textbf{0.97} & \textbf{1.00} & \textbf{0.97} & \textbf{0.99} \\
ROBIN                                        & 23.55 & 0.75 & 0.87 & 27.55 & 0.13 & 0.98 & 0.99 & \textbf{1.00} & \textbf{1.00} & \textbf{0.97} & 0.99 & 0.93 & 0.98 \\
Zodiac                                       & 25.53 & 0.93 & 0.97 & 13.44 & 0.04 & 0.98 & \textbf{1.00} & 0.94 & \textbf{1.00} & \textbf{0.97} & 1.00 & 0.45 & 0.89 \\
 \textbf{Ours}           & 28.18 & 0.94 & 0.97 & 11.32 & 0.03 & \textbf{1.00} & \textbf{1.00} & 0.98 & \textbf{1.00} & 0.96 & \textbf{1.00} & \textbf{0.97} & \textbf{0.99} \\ 
\hline
\multicolumn{14}{c}{\textit{MS-COCO}} \\ 
\hline
DwtDct                                       & \textbf{40.83} & 0.98 &\textbf{0.99} & \textbf{1.22}  & \textbf{0.01} & 0.89 & 0.49 & 0.49 & 0.48 & 0.49 & 0.46 & 0.46 & 0.48 \\
DwtDctSvd                                    & 40.10 & \textbf{0.99} & \textbf{0.99} & 3.19  & \textbf{0.01} & \textbf{1.00} & 0.52 & 0.51 & 0.78 & 0.57 & 0.49 & 0.45 & 0.55 \\
RivaGAN                                      & 39.77 & 0.98 & \textbf{0.99} & 4.81  & 0.02 & \textbf{1.00} & 0.84 & 0.98 & 0.85 & 0.70 & 0.86 & 0.40 & 0.77 \\
StegaStamp                                   & 27.92 & 0.91 & 0.95 & 18.45 & 0.03 & \textbf{1.00} & \textbf{1.00} & 0.61 & 0.94 & 0.76 & 0.88 & 0.44 & 0.77 \\ \hline
Tree-ring                                    & 12.66 & 0.48 & 0.51 & 43.76 & 0.44 & \textbf{1.00} & \textbf{1.00} & 0.99 & \textbf{1.00} & 0.97 & 0.99 & \textbf{0.94} & \textbf{0.98} \\
ROBIN                                        & 22.33 & 0.75 & 0.87 & 20.14 & 0.12 & \textbf{1.00} & 0.99 & \textbf{1.00} & \textbf{1.00} & \textbf{0.98} & 0.97 & \textbf{0.94} & \textbf{0.98} \\
Zodiac                                       & 23.95 & 0.86 & 0.95 & 16.94 & 0.08 & \textbf{1.00} & 0.99 & \textbf{1.00} & \textbf{1.00} & 0.97 & \textbf{1.00} & 0.59 & 0.92 \\
 Ours                    & 27.38 & 0.90 & 0.97 & 7.96  & 0.04 & \textbf{1.00} & \textbf{1.00} & 0.98 & \textbf{1.00} & 0.96 & 0.99 & \textbf{0.94} & \textbf{0.98} \\ \hline
\end{tabular}%
}
\end{table*}

\section{Experimental Evaluation}
In this section, 
we first describe our experimental setup~(Section~\ref{sec:exp_setup}). We then conduct both quantitative~(Section~\ref{sec:quantitative}) and qualitative~(Section~\ref{sec:qualitative}) evaluation of the proposed PT-Mark.
Finally, we present ablation studies in Section~\ref{sec:abla} to analyze the effectiveness of key design components within PT-Mark.
\subsection{Experimental Settings}
\label{sec:exp_setup}
\noindent\textbf{Datasets.}
For fair comparison, we follow the previous work~\cite{wen2023tree, huang2024robin, zhang2024attack} to utilize two mainstream benchmark as our dataset: 
(i).~MS-COCO~\cite{lin2014microsoft}: A large-scale dataset containing approximately 123000 real-world images, each paired with human-annotated captions.  In our experiments, we use 5000 captions from the validation set as prompts to generate corresponding images for watermark evaluation. 
(ii).~DiffusionDB~\cite{wang2022diffusiondb}: This dataset comprises over 2 million text-to-image generation records created using Stable Diffusion models, with each record containing a prompt, the generated image, and associated metadata. Unlike MS-COCO, DiffusionDB focuses exclusively on AI-generated images conditioned on natural language prompts. We randomly sample 8000 prompts to generate images for evaluating watermarking performance.

\vspace{0.3em}
\noindent\textbf{Baselines}. 
To validate the effectiveness of the proposed PT-Mark, we compare it against seven mainstream watermarking methods as baselines:
~(i).Frequency-based Watermarking: DwtDct and DwtDctSvd~\cite{cox2007digital}, which employ frequency decomposition as the embedding strategy. 
~(ii).GAN-based Watermarking: RivaGAN~\cite{zhang2019robust} and StegaStamp~\cite{tancik2020stegastamp}, which embed watermarks through generative adversarial learning. 
~(iii).Diffusion-based Watermarking: Tree-Ring~\cite{wen2023tree}, ROBIN~\cite{huang2024robin}, and Zodiac~\cite{zhang2024attack}, which embed watermarks in the latent space of diffusion models.

\vspace{0.3em}
\noindent\textbf{Evaluation Metrics.}
To comprehensively evaluate our watermarking method PT-Mark, we follow the previous work and 
employ widely-used metrics from \textbf{semantic maintenance} and \textbf{watermark preservation}. 
For semantic maintenance, we utilize typical pixel-level metrics including Peak Signal-to-Noise Ratio~(PSNR), Structural Similarity Index~(SSIM)~\cite{wang2004image}, and Multiscale SSIM~(MSSIM), where higher values indicating better preservation of original image quality. 
To further evaluate the perceptual quality, which better reflects human judgment, we use Learned Perceptual Image Patch Similarity~(LPIPS)~\cite{zhang2018unreasonable} and Fréchet Inception Distance~(FID)~\cite{heusel2017gans} as metrics, with lower scores on these metrics indicate more natural-looking results.
Additionally, we evaluate the robustness of PT-Mark under six common image perturbations: JPEG compression~(quality factor 25), random image cropping~(75\%), Gaussian blur~(radius 4), Gaussian noise~(10\% intensity), brightness adjustments~(color jitter with a brightness factor of 6), and random rotation~(up to 75 degrees).
In all our experiments, we evaluate watermark effectiveness through Area Under the ROC Curve~(AUC) to quantify detection accuracy between watermarked and clean images. 


\vspace{0.3em}
\noindent\textbf{Implementation Details.} In our experiments, we utilize the Stable-Diffusion-v2.1-base model~\cite{rombach2022high} with 50 denoising steps, and adopt the second-order multistep DPM-Solver sampling algorithm. The classifier-free guidance scale is set to 7.5, following the default configuration commonly used in prior works. The generated image size is 512 $\times$ 512.
We employ the Tree-Ring watermark pattern in our experiments, embedding a randomly initialized ring pattern in the last channel of the latent representation, with the radius set to 10.
To guide the optimization of the null-text embedding, we employ Segment Anything Model~(SAM)~\cite{kirillov2023segment} as the pretrained segmentation network for computing the watermark-relevant saliency region.
At each diffusion denoising step, we optimize the null-text embedding 10 times to align the generation trajectory of the watermarked image closely with that of the original image. 
We set the loss weight for semantic maintenance~(i.e., $\lambda_{1}$) to 1.50, and the loss weight for watermark preservation~($\lambda_{2}$) to 0.0007.

\subsection{Quantitative Evaluation}
\label{sec:quantitative}

\textbf{Semantic Preservation}. 
To evaluate the semantic preservation performance of PT-Mark, we assess a series of image quality metrics that reflect the semantic differences between the original image~(without watermark) and the watermarked image generated by PT-Mark, as shown in Table~\ref{tab:quantitative}. 
Overall, PT-Mark consistently outperforms state-of-the-art diffusion-based watermarking methods across all semantic preservation metrics. Notably, the widely used Tree-Ring method significantly degrades image quality, with PSNR dropping to 15.18 and SSIM to 0.56, indicating substantial semantic distortion in both the DiffusionDB and MS-COCO datasets. In contrast, PT-Mark yields over a 20\% improvement in these metrics.
In particular, PT-Mark demonstrates substantial improvements in FID and LPIPS, two metrics closely aligned with human visual perception. 
It achieves FID values below 10 and LPIPS scores below 0.05, outperforming other latent diffusion-based watermarking baselines by a significant margin. 
We also observe that PT-Mark exhibits consistent semantic preservation across different datasets, with only slight variations. In contrast, Zodiac exhibits a significant decline in image quality when the text prompt is altered, indicating that existing diffusion-based watermarking methods may be sensitive to prompt variation and lack robustness in maintaining original semantics.
Compared to traditional methods such as DwtDct and DwtDctSvd, which achieve high image quality~(e.g., PSNR > 38, SSIM > 0.97), PT-Mark achieves comparable perceptual performance~(e.g., SSIM = 0.92, PSNR = 28.18, FID = 11.32 on DiffusionDB), while maintaining stronger watermark traceability.


\vspace{0.3em}
\noindent\textbf{Watermark Robustness}. 
As images are more easily disrupted in real-world or adversarial scenarios, such as when shared on social networks, it is crucial to evaluate the robustness of image watermarking methods. Table~\ref{tab:quantitative} presents the effectiveness of watermark verification under several common real-world perturbations.
We observe that traditional watermarking methods~(shown in the upper half of each table) exhibit poor robustness when exposed to such perturbations, resulting in undetectable watermarks with AUC values dropping below 50\%.
In contrast, PT-Mark demonstrates remarkable robustness across all perturbations, achieving an average AUC of 0.99, while simultaneously maintaining good semantic preservation performance.
For adversarial training-based methods such as RivaGAN and StegaStamp, we find their robustness is limited, particularly under rotation attacks. Although Zodiac achieves relatively good image quality~(e.g., PSNR = 25.53, SSIM = 0.93), it fails to preserve watermark detectability under rotation, leading to a decrease in AUC to 0.89.
We also observe that rotation and Gaussian noise are the most damaging perturbations to watermark traceability. More than half of the evaluated methods exhibit over a 20\% drop in detection AUC under these conditions.
PT-Mark, on the other hand, maintains exceptional resilience, consistently achieving an average AUC of 0.99 across all perturbation types.
Furthermore, we provide results on real-world adaptive watermark removal attacks to assess PT-Mark’s robustness in more challenging scenarios, as detailed in Appendix~\ref{sec:adaptive_attack}.

\vspace{0.3em}
\noindent\textbf{Efficiency}. 
To evaluate the efficiency of PT-Mark, we measure its time cost for generating a single image and compare it with state-of-the-art latent diffusion-based watermarking methods, as shown in Table~\ref{tab:time}. The time cost is categorized into two components: training cost and inference cost. Among the evaluated methods, ROBIN requires training a specific watermark pattern for each user, resulting in significant training overhead. In contrast, the other methods, including Tree-Ring, Zodiac, and PT-Mark, primarily incur time costs during inference and do not require additional training. Although ROBIN achieves a faster inference process than Tree-Ring, it incurs a high training cost, making it several times more time-consuming overall compared to the other methods. 
PT-Mark achieves a good balance between efficiency and performance. 
For instance, PT-Mark improves efficiency with a 4$\times$ reduction in inference time compared to Zodiac. 
Although PT-Mark's inference time is approximately 10 times higher than Tree-Ring's, it significantly outperforms Tree-Ring in terms of semantic preservation. In contrast, while Tree-Ring is lightweight and easy to implement, it suffers from significant degradation in semantic quality.



\begin{table}[h!]
\caption{The time cost of different watermarking methods.}
\centering
\resizebox{\columnwidth}{!}{
\begin{tabular}{l|c|c|c}
\hline
\textbf{Method} & \textbf{Training Cost~(s)} & \textbf{Inference Cost~(s)} & \textbf{Total Cost~(s)} \\
\hline
Tree-ring & 0.00 & 11.65 & \textbf{11.65} \\
ROBIN & 1370.48 & \textbf{3.74} & 1374.21 \\
Zodiac & 0.00 & 684.67 & 684.67 \\
Ours & 0.00 & 149.94 & 149.94 \\
\hline
\end{tabular}
}

\label{tab:time}
\end{table}

\subsection{Qualitative Evaluation}
\label{sec:qualitative}
\begin{figure}[t]
    \centering
    \includegraphics[width=\linewidth]{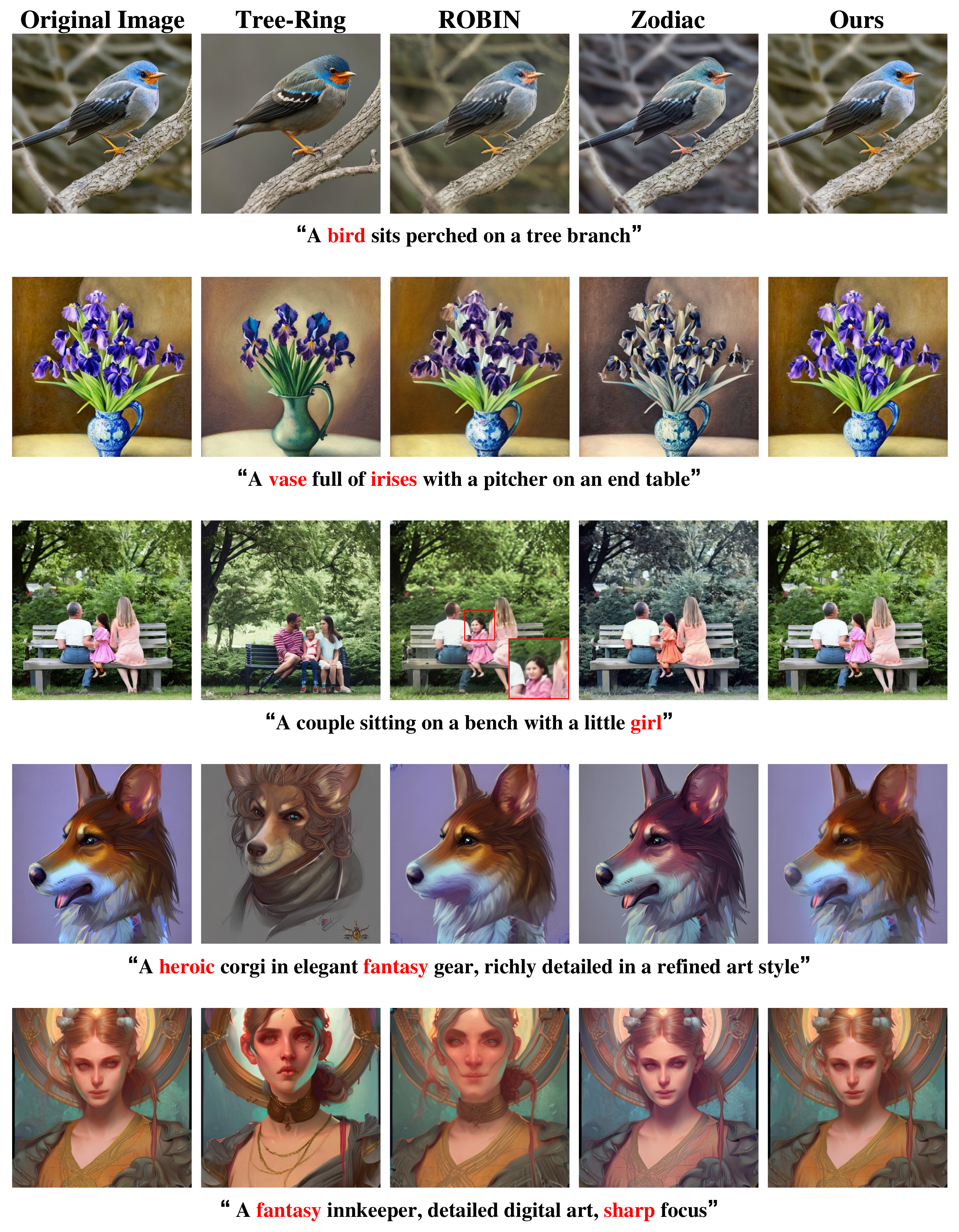}
    \caption{Qualitative evaluation of PT-Mark compared to baseline methods on the MS-COCO dataset~(upper three rows) and Diffusion DB dataset~(lower two rows).}
    \label{fig:compare}
\end{figure}

Considering that traditional watermarking methods fail to withstand real-world perturbations and are therefore impractical for deployment in real-world scenarios, our qualitative evaluation focuses on advanced diffusion-based watermarking methods, including Tree-Ring, ROBIN, and Zodiac.
In Figure~\ref{fig:compare}, we present five representative examples drawn from both the MS-COCO dataset~(upper three rows) and the DiffusionDB dataset~(lower two rows), using the same watermark patterns across all methods for fair comparison. These cases cover a variety of image styles and drawing objects. For enhanced clarity and visual comparison, we recommend viewing the figure in color and zooming in for finer details. 


Overall, PT-Mark demonstrates superior text alignment, semantic fidelity, and visual quality compared to existing diffusion-based watermarking methods. For natural scene images~(e.g., in the first and third rows), PT-Mark preserves finer details and maintains natural color tones. In contrast, Zodiac often introduces unnatural color shifts, while Tree-Ring frequently results in significant semantic drift. Although ROBIN retains similar semantics, it struggles with fine-grained details, such as misaligning the facial orientation of the girl in the third row. For images in a drawing style, PT-Mark consistently outperforms other methods in texture coherence, color consistency, and semantic alignment. Notably, Tree-Ring significantly degrades the artistic expressiveness of the generated content. In the fifth-row example, it causes severe alterations to facial attributes in a digital portrait, failing to capture the ``heroic'' essence of a corgi in fantasy armor, and often generates incoherent, abstract patterns due to disrupted latent distributions. ROBIN preserves global structure but suffers from local distortions, while Zodiac produces dim, desaturated colors across all samples.
In contrast, PT-Mark generates watermarked images that are visually indistinguishable from their clean, unwatermarked counterparts, preserving both the structure and style of the original generation.

\begin{table*}[]
\centering
\caption{Ablation study of PT-Mark with different modules. The ablation settings are elaborated in Section~\ref{sec:abla}.}
\label{tab:abla}
\resizebox{\textwidth}{!}{%
\begin{tabular}{l|ccc|ccccc|cccccccc}
\hline
\multicolumn{1}{c|}{\multirow{2}{*}{Method}} &
  \multicolumn{3}{c|}{Module} &
  \multicolumn{5}{c|}{Image Quality} &
  \multicolumn{7}{c}{Watermarking Robustness} &
   \\
\multicolumn{1}{c|}{} &
  WE &
  SA &
  WP &
  PSNR $\uparrow$ &
  SSIM $\uparrow$ &
  MSSIM $\uparrow$ &
  FID $\downarrow$ &
  LPIPS $\downarrow$ &
  Clean &
  JPEG &
  Crop &
  Blur &
  Noise &
  Bright &
  Rotation &
  Avg \\ \hline
\textcolor{gray!80}{Clean Image} &
   \textcolor{gray!80}{-} &
   \textcolor{gray!80}{-} &
   \textcolor{gray!80}{-} &
  \textcolor{gray!80}{$\infty$} &
  \textcolor{gray!80}{1.00} &
  \textcolor{gray!80}{1.00} &
  \textcolor{gray!80}{0.00} &
  \textcolor{gray!80}{0.00} &
  \textcolor{gray!80}{-} &
  \textcolor{gray!80}{-} &
  \textcolor{gray!80}{-} &
  \textcolor{gray!80}{-} &
  \textcolor{gray!80}{-} &
  \textcolor{gray!80}{-} &
  \textcolor{gray!80}{-} &
  \textcolor{gray!80}{-} \\
Tree-Ring &
  $\checkmark$ &
   &
   &
  15.18 &
  0.56 &
  0.59 &
  42.97 &
  0.37 &
  \textbf{1.00} &
  \textbf{1.00} &
  \textbf{1.00} &
  \textbf{1.00} &
  \textbf{0.97} &
  \textbf{1.00} &
  \textbf{0.97} &
  \textbf{0.99} \\
PT-Mark~(w/o WP) &
  $\checkmark$ &
  $\checkmark$ &
   &
  \textbf{28.33} &
  \textbf{0.94} &
  \textbf{0.97} &
  \textbf{10.73} &
  \textbf{0.03} &
  \textbf{1.00} &
  0.92 &
  0.92 &
  0.95 &
  0.77 &
  0.87 &
  0.80 &
  0.87 \\
PT-Mark &
  $\checkmark$ &
  $\checkmark$ &
  $\checkmark$ &
  28.18 &
  \textbf{0.94} &
  \textbf{0.97} &
  11.32 &
  \textbf{0.03} &
  \textbf{1.00} &
  \textbf{1.00} &
  0.98 &
  \textbf{1.00} &
  0.96 &
  \textbf{1.00} &
  \textbf{0.97} &
  \textbf{0.99} \\ \hline
\end{tabular}%
}
\end{table*}

\subsection{Ablation Study}
\label{sec:abla}

In this section, we further validate the effectiveness of each component and hyperparameter in PT-Mark through a comprehensive ablation study. Our experiments on module effectiveness focus on two core components: Semantic Alignment~(SA), which optimizes the null-text embedding to ensure the edited latent state remains close to the original generation trajectory, and Watermark Preservation~(WP), which refines the null-text embedding to align the edited latent with the watermarked trajectory in salient regions. For reference, we also include results from a baseline configuration without watermark embedding~(w/o WE) to isolate the impact of watermark-related modules. For the hyperparameter analysis, we focus on evaluating the effect of the number of null-text embedding optimization iterations, and conduct an ablation study to explore the impact of the starting timestep for pivotal tuning.


\vspace{0.3em}
\noindent\textbf{Module Effectiveness.}
The results of the evaluation on module effectiveness are shown in Table~\ref{tab:abla}. We observe that watermark integration significantly degrades image quality, 
with Tree-Ring yielding an SSIM of 0.56, a PSNR of 15.18, and an FID of 42.97.
We further demonstrate that applying pivotal tuning only on the original trajectory~(Semantic PT) substantially improves visual quality, with SSIM rising to 0.94, PSNR increasing to 28.33, and FID decreasing to 10.73. Additionally, LPIPS significantly decreases to 0.03.
However, this optimization comes at the cost of robustness, particularly under transformations such as noise~(0.77), brightness~(0.87), and rotation~(0.80), which reduces the average robustness to 0.87.
By editing the entire diffusion denoising process with the guidance of the watermark trajectory~(PT-Mark), we mitigate the side effects introduced by semantic pivotal tuning, improving robustness while still maintaining high image quality compared to applying only semantic pivotal tuning.

\begin{figure}[t]
    \centering
    \includegraphics[width=\linewidth]{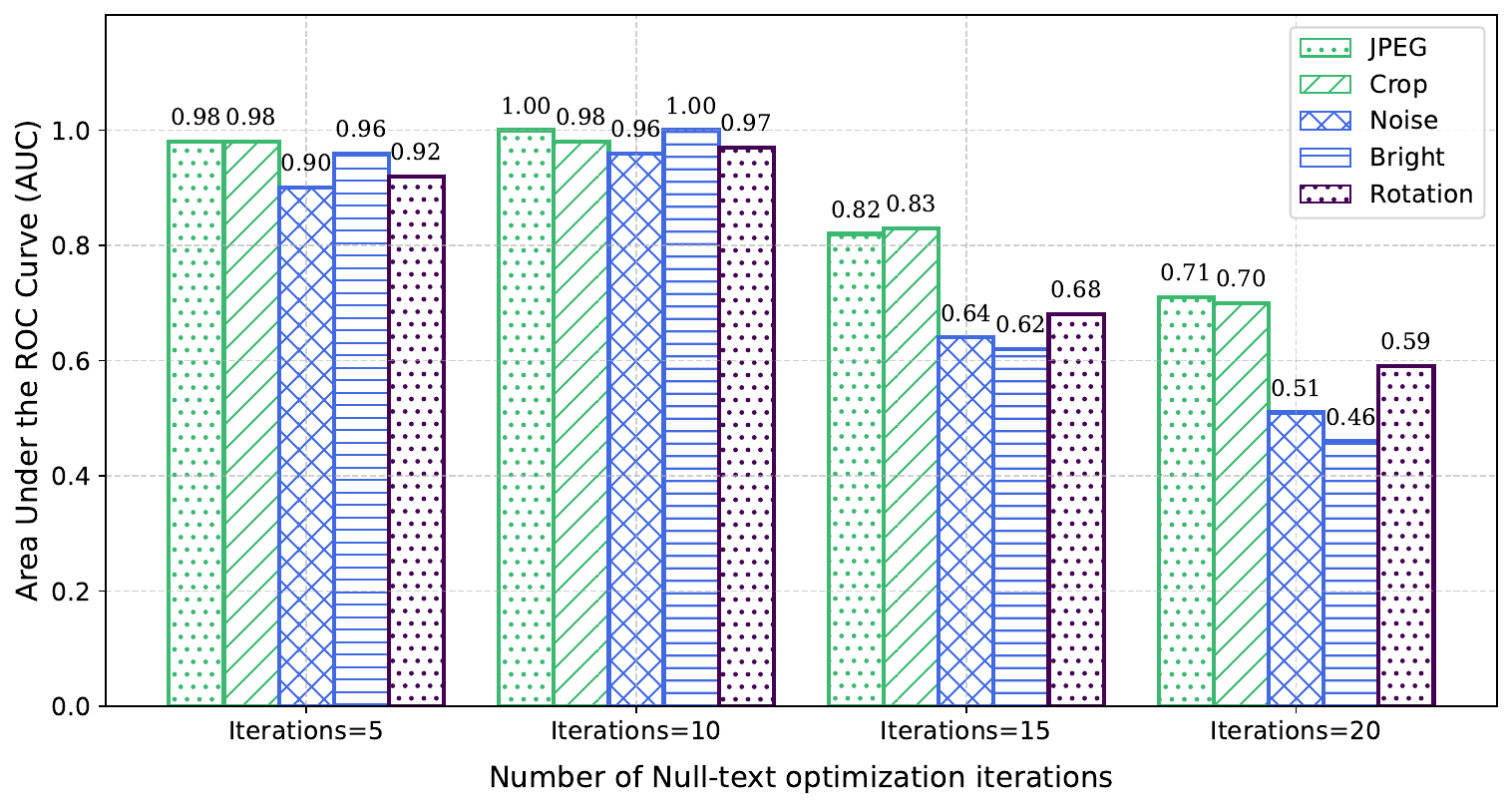}
    \caption{Impact of Null-text optimization iterations on AUC under different degradations.}
    \label{fig:ablation_distortion}
\end{figure}

\begin{figure}[t]
    \centering
    \includegraphics[width=\linewidth]{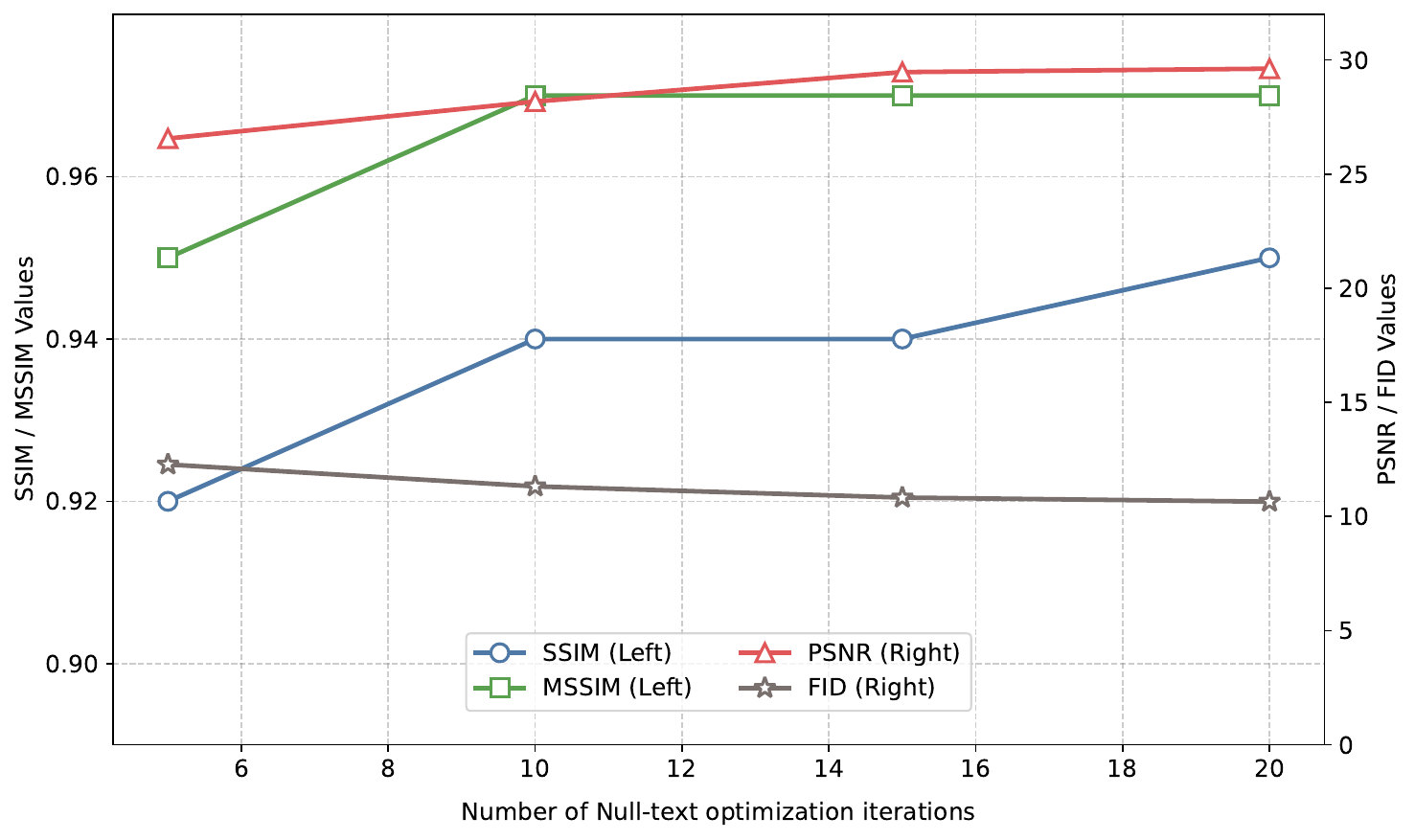}
    \caption{Impact of Null-text optimization iterations on image quality metrics.}
    \label{fig:ablation_quality}
\end{figure}

\vspace{0.3em}
\noindent\textbf{Number of Null-text optimization iterations.}
We conduct an ablation study to investigate the impact of the number of null-text optimization iterations in the trajectory optimization process used during image watermark embedding. Specifically, we evaluate how varying the number of iterations~(5, 10, 15, 20) affects both the perceptual quality of the watermarked images and their robustness against common real-world perturbations. As shown in Figure~\ref{fig:ablation_quality}, increasing the number of iterations generally enhances the visual quality of the generated watermarked images. Metrics such as PSNR and SSIM steadily improve, FID decreases from 12.28 to 10.65, and LPIPS reaches its optimal value at Step = 10, indicating better fidelity and perceptual similarity to the original image.
However, this improvement comes at the cost of robustness. As illustrated in Figure~\ref{fig:ablation_distortion}, optimization with higher iterations~(15 or 20) results in significantly reduced classification accuracy under common image-level distortions, such as cropping, noise, blur, brightness changes, and rotation. For example, under a brightness attack, accuracy drops from 1.00 at Step = 10 to just 0.46 at Step = 20, suggesting that too much optimization on semantics may lead to overfitting to clean data distributions, thereby making the embedded watermark more vulnerable to real-world perturbations.
Overall, step = 10 offers the best trade-off, achieving high watermarked image quality while maintaining strong resilience to various image perturbations. 

\begin{figure}[t]
    \centering
    \includegraphics[width=\linewidth]{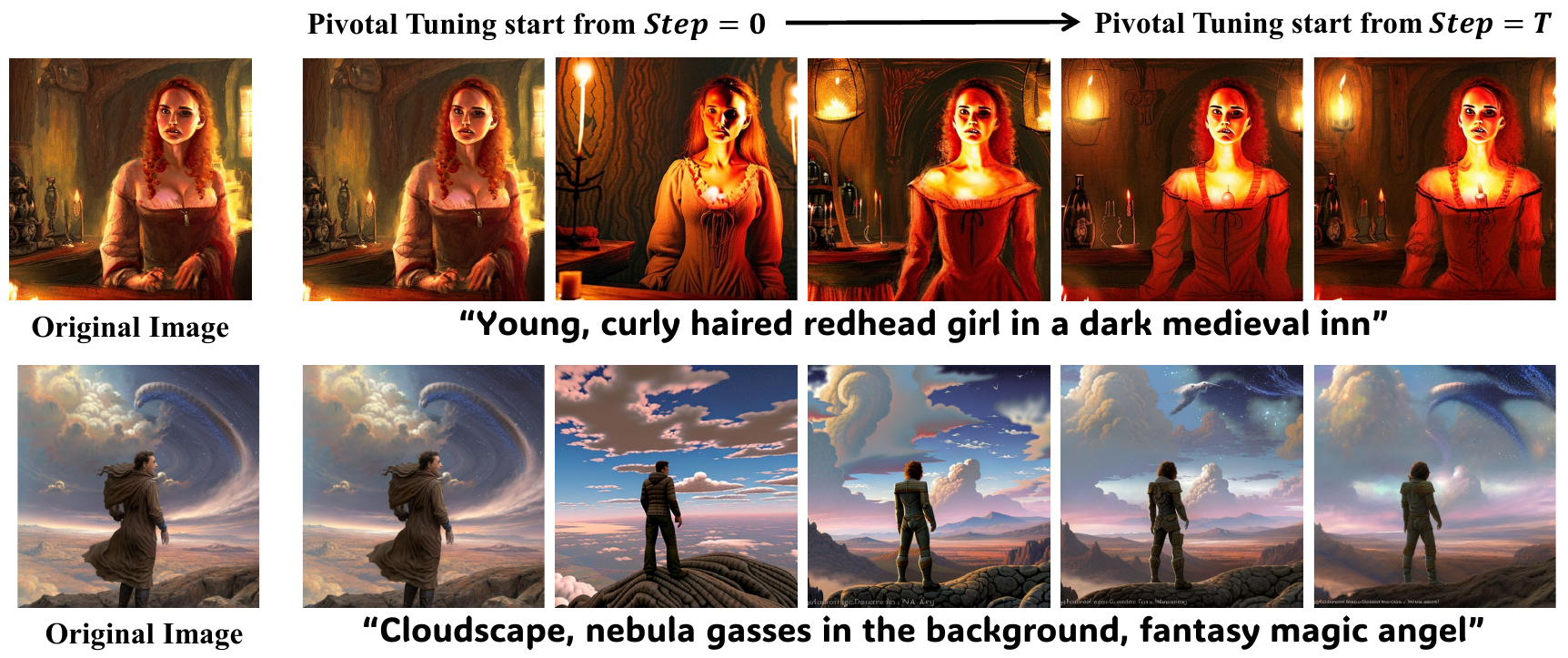}
    \caption{Impact of pivotal tuning step on image quality.}
    \label{fig:starting_timestamp}
\end{figure}

\vspace{0.3em}
\noindent\textbf{Starting Step of Pivotal Tuning.}
To investigate the influence of the starting step of pivotal tuning, we present examples of generated watermarked images edited from different steps, as shown in Figure~\ref{fig:starting_timestamp}.
We observe that starting the tuning process from $Step=0$ results in images with the highest performance in preserving the original semantics, especially in the intricate details of the image, such as facial features, lighting, and textures.
This early intervention allows the optimization process to guide the generation trajectory from the beginning, ensuring that both image quality and semantic consistency align well with the original image.
In contrast, employing pivotal tuning at later steps results in noticeable degradation in semantic preservation. Specifically, we observe that the generated images increasingly drift from the original image semantics as the tuning occurs later in the process.

%% file: 6_conclusion.tex
\section{Conclusion}
In this paper, we proposed a novel invisible watermarking method for Text-to-image Diffusion Models, called Semantic-aware Pivotal Tuning Watermark~(PT-Mark). We achieved the goal of semantic preservation after embedding watermarks by editing the entire diffusion denoising process.
By incorporating an additional pivotal tuning branch, PT-Mark optimized a learnable null-text embedding to disentangle semantics from watermark patterns during the diffusion denoising process and generated a steering vector that aligns the entire generation process while preserving the traceability of the watermark.
Moreover, both qualitative and quantitative evaluations demonstrated that our method can embed robust watermarks without disrupting the original semantics of the generated image.